# Mid-Infrared Polarization of the Diffuse Interstellar Medium toward CygOB2-12[*,1]

Charles M. Telesco,[1] Frank Varosi,[1] Christopher Wright,[2] Bruce T. Draine,[3]
Sergio Jose Fernández Acosta,[4] and Christopher Packham[5]

[1]*Department of Astronomy, University of Florida, 211 Bryant Space Science Center, Gainesville, FL 32611, USA*
[2]*School of Science, UNSW Canberra, Canberra BC 2610, Australia*
[3]*Department of Astrophysical Sciences, Princeton University, Princeton, NJ, 08544-1001 USA*
[4]*Gran Telescopio de Canarias, Cuesta de San José s/n, E-38712, Breña Baja, La Palma, Spain*
[5]*Physics and Astronomy Department, University of Texas at San Antonio, San Antonio, TX 78249, USA*

## ABSTRACT

We present the first mid-IR detection of the linear polarization toward the star CygOB2-12, a luminous blue hypergiant that, with $A_V \approx 10$ mag of foreground extinction, is a benchmark in the study of the properties of dust in the diffuse interstellar medium. The 8-13 $\mu$m spectropolarimetry, obtained with the CanariCam multi-mode camera at the Gran Telescopio Canarias (GTC), shows clear trends with wavelength characteristic of silicate grains aligned in the interstellar magnetic field. The maximum polarization, detected with $7.8\sigma$ statistical significance near 10.2 $\mu$m, is $(1.24 \pm 0.28)$ % with position angle $126° \pm 8°$. We comment on these measurements in the context of recent models for the dust composition in the diffuse interstellar medium.

## 1. INTRODUCTION

Silicate dust is a key constituent of the interstellar medium (ISM), intimately participating in its chemistry and energetics. Determining its shapes, sizes, and composition, as well as the nature of its interactions with other ISM constituents, is therefore of broad observational and theoretical interest. Interstellar silicate particles have been studied extensively with mid-IR spectroscopy and polarimetry, with particular focus on the broad 9.7 $\mu$m resonant feature due to the Si-O stretching mode (e.g., Smith et al. 2000). The feature's profile has provided key constraints on both the mineralogy and morphology of the particles in a wide variety of astrophysical environments. Generally, interstellar silicate particles have an amorphous structure, with crystalline silicate signatures being apparent in some environments.

Despite this progress, probing the silicates in the lower-density, diffuse ISM has been challenging due to the large physical distances needed to accrue ISM optical depths sufficient for informative mid-IR spectroscopy and spectropolarimetry. Most stars are too faint to be useful as probes, but at least one sightline has proven fruitful, that toward the luminous blue hypergiant star CygOB2-12. The high luminosity (1.4 × 10$^6$ L$_\odot$, Hensley & Draine 2020) and relative proximity (1.75 kpc, Clark et al. 2012) of CygOB2-12 make it sufficiently bright to be observable across the spectrum through ~10 mag of visual extinction. The lack of ice features and the appearance of the extinction curve itself (Whittet 2015) indicate that this large extinction likely arises in diffuse, rather than dense, clouds. This fortuitous combination of high foreground extinction and high stellar brightness has therefore made CygOB2-12 a unique probe of the diffuse ISM, particularly its silicates.

Mid-IR spectroscopy of CygOB2-12 has provided considerable insight into the detailed composition of the silicate particles (Rieke 1974; Fogarty et al. 2016; Hensley & Draine 2020). Mid-IR spectropolarimetry can provide additional and unique insight into the particle shape, porosity, and dielectric properties (Hildebrand 1988;

---

Draine & Hensley 2021b), but to date no mid-IR spectropolarimetry of CygOB2-12 has been available. Here we present the first detection of the polarization of the 9.7 $\mu$m silicate feature toward CygOB2-12.

## 2. OBSERVATIONS & DATA REDUCTION

We obtained spectropolarimetry of CygOB2-12 on UT 2020 August 5 and October 28 with CanariCam, the mid-IR multi-mode facility camera on the 10.4-m Gran Telescopio CANARIAS (GTC) on La Palma, Spain (Telesco et al. 2003). CanariCam employs a 320×240pixel Raytheon Si:As impurity band conduction (IBC) detector array with a pixel scale of 0".079, which provides a field of view of 26"×19" with Nyquist sampling of the diffraction-limited (~0".3) point-spread function at 8 $\mu$m. Polarimetry is accomplished by inserting into the beam a half-wave plate rotated to angles of 0°, 22.5°, 45°, and 67.5° and a Wollaston prism that separates ordinary (*o*) and extraordinary (*e*) rays, which are recorded by the detector simultaneously. A field mask inserted to prevent overlap of the *o* and *e* rays limits the short dimension of the field of view to 2".56, which, in the spectropolarimetric mode, is the slit length.

Our low-resolution ($\lambda/\Delta\lambda \approx 50$) spectropolarimetry spans the wavelength range 8.1–13.0 $\mu$m and consists of nine individual data sets obtained using a 1".02-wide slit. The total on-source integration time was 2979 s with an equal amount of integration time for sky-reference observations. We used the Cohen standard stars HD188056 and HD213310 (Cohen et al. 1999) for flux and point-spread-function calibration and for telluric correction of the Stokes I spectrum. The standard star AFGL 2591, selected from Smith et al. (2000), was used to calibrate the polarization position angle. The standard mid-IR chop-nod technique was applied with a 5".7 N-S chop throw. CygOB2-12 and the photometric standards were observed at airmasses in the range 1.08 - 1.32 and values of the line-of-sight atmospheric precipitable water vapor (PWV) of ~2.6 mm for the October 28 run. (PWV data are unavailable for the August 5 run.) Observing conditions were generally good, although the widest CanariCam slit was needed to accommodate somewhat elevated seeing.

We extracted one-dimensional spectra of *o* and *e* rays at each half-wave-plate angle by summing pixel fluxes along the slit and computing normalized Stokes parameters $q = Q/I$ and $u = U/I$, where $I$ is the total observed intensity, using the ratio method (Tinbergen 2005) applied to the eight spectra (*o/e* rays and four half-waveplate angles). We assumed the instrumental polarization to be (0.6 ± 0.1) % as measured systematically during the CanariCam commissioning and confirmed during subsequent observations. Note that the correction for instrumental polarization, the orientation of which depends on the telescope pupil orientation during the observation, is applied to $q$ and $u$ before deriving $p$.

The Stokes parameters were co-added into 0.4 $\mu$m-wide bins. Standard deviations of data within each bin provided estimates of the stochastic uncertainties $\sigma_q$ and $\sigma_u$ associated with the average normalized Stokes parameters for each bin. We computed the degree of polarization $p = (q^2 + u^2 - \sigma_p^2)^{0.5}$ in each bin, where the last term (the "debias" term) is introduced to remove a positive offset in the signal floor resulting from the squared standard deviation of the background-noise errors. The debias term, computed using the Modified Asymptotic Estimator method of Plaszczynski et al. (2014), is the standard deviation of the degree of polarization in each bin computed using the propagation of errors formula $\sigma_p = ((\sigma_q q)^2 + (\sigma_u u)^2)^{0.5}/p$. The polarization position angle PA, measured east from north, was computed as $\theta = 0.5 \arctan(u/q)$, with uncertainty $\sigma_\theta = \sigma_p/2p$ (Patat & Romaniello 2006).

## 3. RESULTS

Figure 1 shows our 8-13 $\mu$m intensity (Stokes I) spectrum along with the *Spitzer* IRS spectrum from Hensley & Draine (2020). The central part of our measured spectrum, i.e., between roughly 8 and 11 $\mu$m, has high signal-to-noise, typically ~400, and error bars across that region are negligibly small in this plot. Outside that region, noise due to atmospheric variability and thermal emission is higher and is evident as low-level structure in the spectrum in Fig. 1; those features are unassociated with CygOB2-12. The *Spitzer* fluxes have been multiplied by 1.12 to provide the best fit by eye to our data. The 12% difference in the photometry is consistent with our estimated photometric uncertainty of ±10%. The comparison in Fig. 1 indicates that our observed spectropolarimetry returns a silicate absorption profile that agrees well with previous spectroscopic results, and it links our spectropolarimetry to the analysis by Draine & Hensley (2021a, b) in so far as it is constrained by the silicate-feature absorption profile.

Our 8-13 $\mu$m spectropolarimetry of CygOB2-12 is shown in Fig. 2. The plotted error bars for both $p$ and PA reflect the $1\sigma$ dispersions of the measured values within each 0.4 $\mu$m bin. The noise is largest at the edges of the atmosphere's 10 $\mu$m transmission window and near the 9.8 $\mu$m ozone feature, consistent with the dominance of sky noise over most of the spectrum. Repeated measurements of PA calibrators, particularly AFGL 2591, indicate systematic errors of order ±0.2% in polarization and ±8° in PA. Since each of these uncertainties, along with the ±0.1% associated with the instrumental polarization correction noted earlier, affects all points in the spectropolarimetry the same way, they are not included in the error bars shown in Fig. 2.

Figure 2 reveals a central polarization maximum, with values decreasing toward the shortest and longest wavelengths, and a constant PA value of 126°± 5° (dashed line). The spectrum has the broad profile characteristic of silicates (e.g., Smith et al. 2000), and, given the intensity spectrum (Fig. 1), the constant PA value is consistent with the observed silicate feature resulting from only one process: absorption. For 0.4 $\mu$m binning, the maximum polarization, observed at 10.2 $\mu$m, is (1.24 ± 0.16) %. Including the systematic uncertainty of ±0.22%, the total uncertainty in this value is ±0.28 %. The peak polarization is therefore detected with 7.8$\sigma$ statistical significance, with its value determined with 4.4$\sigma$ statistical significance. Experimenting with various levels of binning, we estimate that the value for the wavelength of maximum polarization is defined to within ±0.1 $\mu$m.

The observed PA is reasonably consistent with PA values observed for CygOB2-12 at other, shorter, wavelengths: 118° in the R-band (Kobulniky et al. 1994) and ~117° between 430 and 500 nm, as compiled from multiple sources by Whittet (2015). Most importantly, we observe no systematic variation of PA with wavelength expected if either emission or scattering were contributing to the polarization.

## 4. DISCUSSION

### 4.1. *Polarization Profile*

In Fig. 3 (left panel), we overlay our spectropolarimetry with that from Wright et al. (2002) who drew on data from the extensive UKIRT/AAT spectropolarimetric survey by Smith et al. (2000). Wright et al. (2002) combined observations from the two Wolf-Rayet (WR) stars WR48a and WR112 (AFGL 2104), which have non-silicate, carbonaceous dust shells and thus should not contribute to any silicate absorption along those sightlines. The silicate spectral and polarimetric features detected in the two WR stars are identical to each other and thought to arise wholly, or nearly so, in the diffuse ISM. We take their composite spectrum as the best available example of the observed diffuse ISM silicate profile. This direct comparison of the two data sets implies that, within the uncertainties, the CanariCam polarization profile of CygOB2-12 is similar to that for the WR stars. Both distributions peak near 10.2 $\mu$m and have full widths of ~2.4 $\mu$m at half the maximum polarization.

In Fig. 3 (right panel), we compare the same normalized CanariCam data to three different profiles. The blue dashed line (diffuse ISM: DISM) is a fit by Wright et al. (2002) to their WR data of a polarization profile calculated for non-porous oblate glassy silicates with b/a = 2; we have slightly renormalized that curve to pass through our maximum binned polarization value. The gray dot-dash line (Astrodust: Ad), from Draine & Hensley (2021a), discussed further below, is their model profile for oblate astrodust spheroids with b/a = 1.6 and porosity 0.2 that is among their best fits to the Wright et al. (2002) data. The dark solid line is the polarization profile for the Becklin-Neugebauer (BN) protostar in Orion, as observed by Smith et al. (2000); we note that Aitken, Smith, & Roche (1989), as well as our team, also observed BN and obtained the same result.

The key conclusion from this comparison is that CygOB2-12 displays a polarization profile consistent with the polarization arising in amorphous silicate particles in the intervening diffuse ISM. Within this broad constraint, though, a range of profile shapes is possible, as Fig. 3 illustrates.

### 4.2. *Magnitude of the Polarization*

Recently, Hensley & Draine (2020) presented a new analysis of *Infrared Space Observatory* and *Spitzer* mid-IR spectroscopy of CygOB2-12 and modeled the star and stellar wind spectra to determine anew the 2.4 to 37 $\mu$m extinction curve for the diffuse ISM toward CygOB2-12. That study was extended by Draine & Hensley (2021a, b) to take advantage of additional information about dust grains accessible with polarimetry of the particles aligned in the interstellar magnetic field. They considered a new model for dust in the diffuse ISM, namely, that the material making up the bulk of the dust is an idealized mixture ("astrodust") of different constituents. Each particle larger than ~0.01 $\mu$m incorporates that mixture and is characterized by its effective dielectric function, shape, and porosity. Silicates make up about half of the mass of these mixed-composition, silicate-bearing particles. Importantly, the models incorporate the assumption that the optical extinction, 10 $\mu$m polarization, and submillimeter polarization arise from the same grains.

Using the totality of data available across the spectrum for CygOB2-12, Draine & Hensley (2021a, b) produced model polarization profiles for astrodust, one of which we show in Fig. 3., and they predict a peak polarization at 10 $\mu$m of (2.1 ± 0.3) %. This is to be compared with our measurement of the maximum polarization near 10.2 $\mu$m of $p$ = (1.24 ± 0.28) %, which includes all known systematic uncertainties in our results. The observed polarization is about half the predicted value, but statistically the difference is at the level of a few standard deviations, depending on the details of the systematic uncertainties.

To be clear, the principal assumptions of the astrodust models are as follows:

1. The determination by Hensley & Draine (2020) of the mid-IR extinction to CygOB2-12 is correct;
2. The dust on the sightline to CygOB2-12 has optical properties similar to the dust at intermediate and high Galactic latitudes;
3. The grains are approximated by partially aligned spheroids;
4. A single type of dust dominates the opacity from the visible to the submillimeter.

Should the difference between predicted and observed 10 $\mu$m polarization be confirmed by new observations or further analysis, then one or more of the assumptions underlying the astrodust models would need to be reconsidered.

For example, per assumption 3, the astrodust model has thus far considered only spheroids (both prolate and oblate), with allowed shapes limited to those that are consistent with the observed magnitude of the starlight optical polarization as well as its ratio with the observed submillimeter polarization; each shape then determines a ratio of 10 $\mu$m polarization to optical starlight polarization. Real interstellar grain shapes may differ substantially from perfect spheroids, but it is not yet known how the ratio of starlight optical polarization (wavelengths comparable to grain size) to 10 $\mu$m polarization (wavelengths large compared to grain size) varies for other grain shapes. It is possible that non-spheroidal grain shapes lower the predicted ratio of 10 $\mu$m polarization to starlight polarization. Calculations to explore this are in progress using the discrete dipole approximation (Draine & Flatau 1994).

In addition, the dielectric function used in the astrodust model (Draine & Hensley 2021a) is derived from the observed wavelength-dependent infrared extinction inferred from spectrophotometry of Cyg OB2-12 (Hensley & Draine 2020). A recent study (Gordon et al. 2021) finds the 5-20 $\mu$m extinction to be significantly lower than the extinction law found by Hensley & Draine (2020); the silicate feature strength is in agreement, but the underlying continuum extinction is lower. If Gordon et al. are correct, then the derived dielectric function would be expected to lower the peak polarization at 10 $\mu$m. In any case, it is clear that the 10 $\mu$m polarization on diffuse ISM sightlines provides a strong constraint on interstellar grain models.

## 5. CONCLUSIONS

Using CanariCam at the GTC, we have made the first detection of mid-IR polarization toward CygOB212. Our spectropolarimetric observations reveal an 8-13 $\mu$m profile with a maximum near (10.2 ± 0.1) $\mu$m, decreasing to longer and shorter wavelengths consistent with that expected for amorphous silicates aligned in the interstellar magnetic field. Given the known properties of this sightline, this profile is attributed to silicate-bearing particles in the diffuse ISM. The maximum value of the detected polarization is 1.24 % at 10.2 $\mu$m. The total uncertainty in this value is 0.28 %, of which 0.16 % is attributed to measurement fluctuations and 0.22 % to systematics including 0.1 % to the correction for instrumental polarization. This value differs from that predicted for the recently proposed astrodust model for the diffuse ISM and may motivate reconsideration of some of its underlying assumptions.

## ACKNOWLEDGEMENTS

We acknowledge the outstanding support of the GTC science and engineering staff who made these observations possible. This research was supported in part by the National Science Foundation under grants AST1515331 and AST-1908625 to CMT and AST-1908123 to BTD. CMW acknowledges financial support during the period 2012-2017 from an Australian Research Council Future Fellowship FT100100495. The upgrade of CanariCam was co-financed by the European Regional Development Fund (ERDF), within the framework of the "Programa Operativo de Crecimiento Inteligente 20142020", project "Mejora de la ICTS Gran Telescopio CANARIAS (2016-2020)."

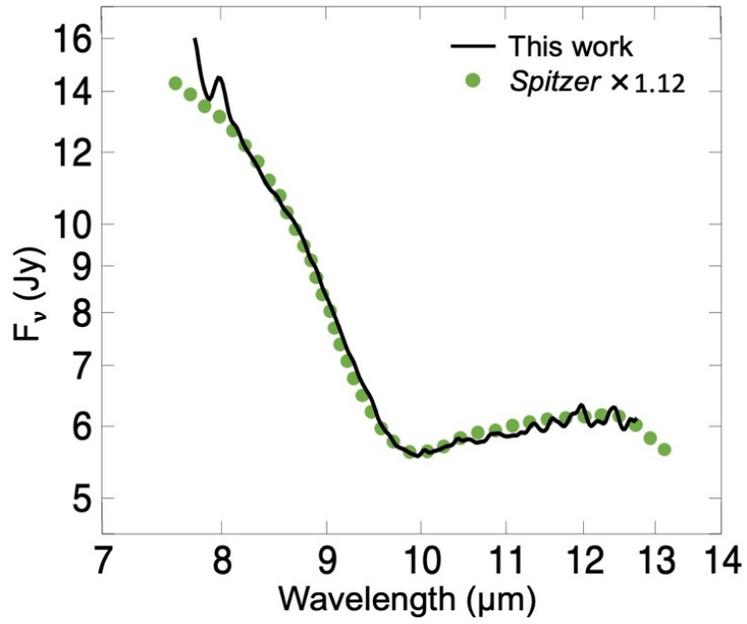

**Figure 1.** Comparison of CanariCam 8-13 µm intensity (Stokes I) spectrum of CygOB2-12 with *Spitzer* IRS spectrum spanning the silicate absorption feature. The *Spitzer* fluxes have been multiplied by 1.12 to provide the best fit by eye to the CanariCam spectrum. Weak spectral features beyond ~10.3 µm and shortward of ~8.5 µm are noise.

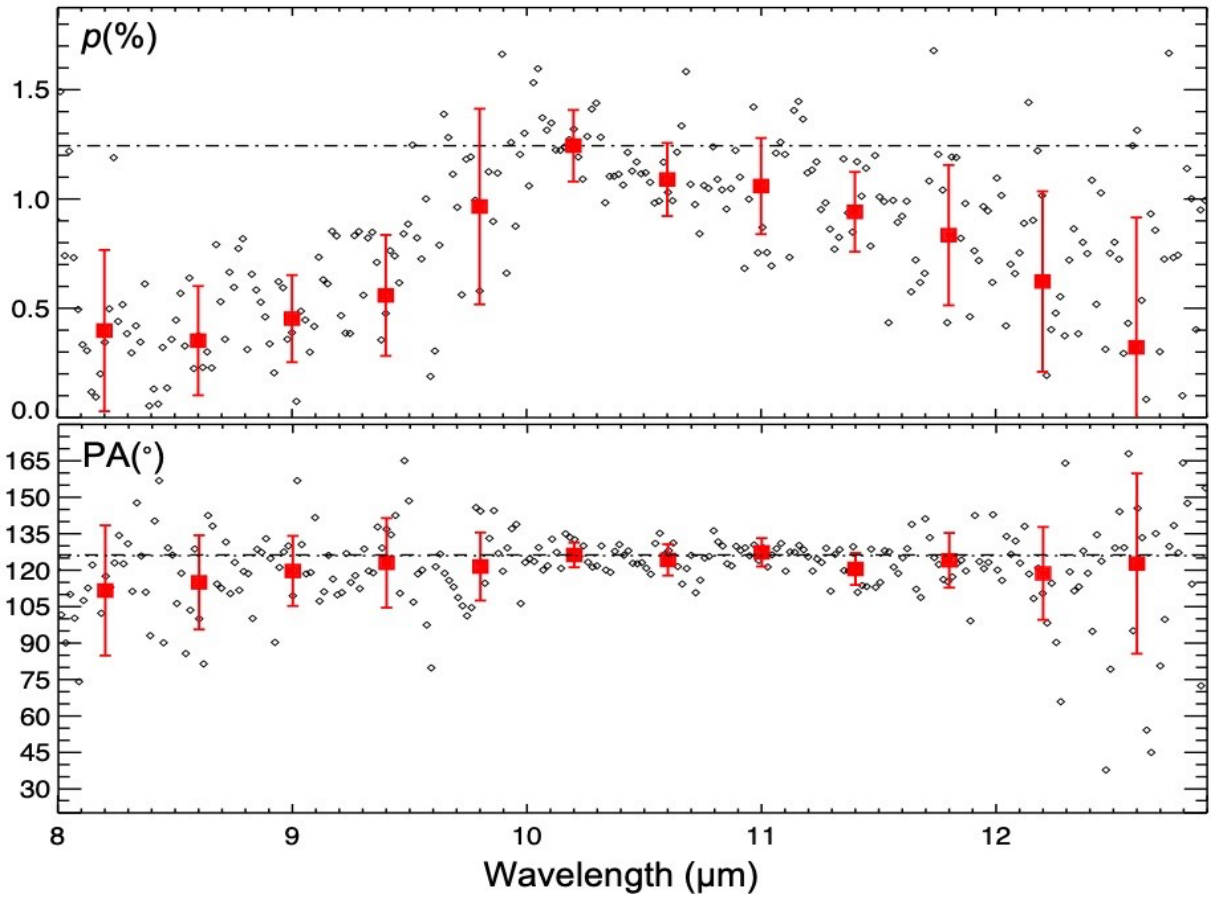

**Figure 2.** CanariCam 8-13 μm spectropolarimetry of CygOB2-12. Panels show percentage polarization *p* and position angle PA measured east of north. Small diamonds are fully-corrected unbinned data. Square symbols are values with 0.4 μm binning. Error bars are 1σ deviations for corresponding binned subsets and do not include estimated systematic uncertainties. Dot-dashed horizonal lines *p* = 1.24% and PA = 126°. The slit width of 1″.04 provided a spectral resolving power of ∼50 and a resolution of ∼0.2 μm near 10 μm.

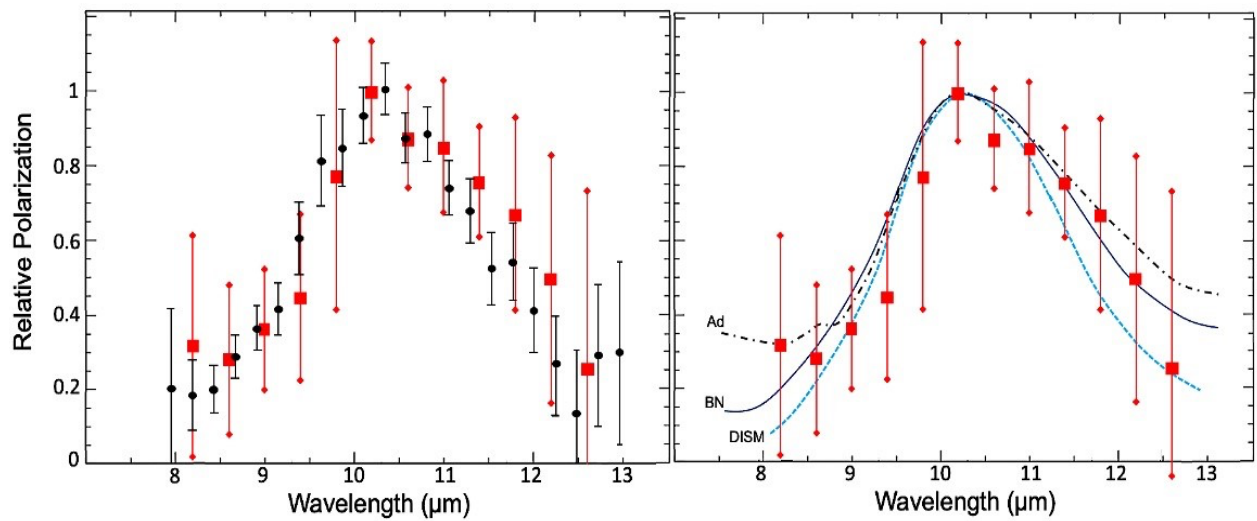

**Figure 3.** Comparison of 0.4 μm-binned and normalized CanariCam polarization data (red squares) of CygOB2-12 to: (Left) previous spectropolarimetry of WR48a and WR112/AFGL 2104 (filled circles) by Smith et al. (2000) as presented in Wright et al. (2002) for diffuse ISM (DISM); (Right) Astrodust model (dot-dashed line, Ad) by Draine & Hensley (2021a) for oblate spheroids with b/a = 1.6 and porosity 0.2; DISM model (dashed line) by Wright et al. (2002) calculated for non-porous oblate glassy silicates with b/a = 2; observed BN profile fit (solid line) to spectropolarimetry of Becklin-Neugebauer protostar by Smith et al. (2000). Peaks for curves and data sets are normalized to unity.